# 5G Networks and IoT Devices: Mitigating DDoS Attacks with Deep Learning Techniques

Reem M. Alzhrani and Mohammed A. Alliheedi
Department of Computer Science, Al-Baha University, Al-Baha, Saudi Arabia
rrmsh.92@gmail.com , malliheedi@bu.edu.sa

***Abstract*— The development and implementation of Internet of Things (IoT) devices have been accelerated dramatically in recent years. As a result, a super-network is required to handle the massive volumes of data collected and transmitted to these devices. Fifth generation (5G) technology is a new, comprehensive wireless technology that has the potential to be the primary enabling technology for the IoT. The rapid spread of IoT devices can encounter many security limits and concerns. As a result, new and serious security and privacy risks have emerged. Attackers use IoT devices to launch massive attacks; one of the most famous is the Distributed Denial of Service (DDoS) attack. Deep Learning techniques have proven their effectiveness in detecting and mitigating DDoS attacks. In this paper, we applied two Deep Learning algorithms Convolutional Neural Network (CNN) and Feed Forward Neural Network (FNN) in dataset was specifically designed for IoT devices within 5G networks. We constructed the 5G network infrastructure using OMNeT++ with the INET and Simu5G frameworks. The dataset encompasses both normal network traffic and DDoS attacks. The Deep Learning algorithms, CNN and FNN, showed impressive accuracy levels, both reaching 99%. These results underscore the potential of Deep Learning to enhance the security of IoT devices within 5G networks**.

***Keywords: DDoS Attack, IoT, 5G Network, OMNeT++, Deep Learning*.**

## *I. INTRODUCTION*

IoT has transformed the way we live, communicate, play, and work. IoT is one of the enabling technologies for 5G, allowing the coexistence of different technologies. Important issues for IoT-based 5G networks include the need for high data rates, low latency, and effective spectrum use [1] [2]. IoT applications have grown in popularity in a variety of fields, including smart homes, e-health, smart cities, and smart connected devices [3]. By 2025, the number of IoT devices is expected to exceed 30 billion [4]. IoT applications are expected to make up a significant portion of the 41 million 5G connections globally by 2024, making them a prominent use case for 5G networks in the future [5]. As a result, the IoT requires a super-network infrastructure to manage and govern massive volumes of data. 5G technology emerges as a comprehensive wireless solution, offering significant advantages such as greater network capacity, low latency, high integrity, higher spectral efficiency, and greater bandwidth than any previous generation [6]. IoT devices are vulnerable to attacks due to their widespread use. The fast growth of IoT applications raises security concerns, and attackers target the vulnerabilities in these devices [3] [7]. The most common attacks that can infect IoT devices are DDoS, Denial of Service (DoS), Data Leakage, Malicious Code Injection, Routing Attacks, Data Transit Attacks, etc. [8]. According to Cloudflare's data, DDoS attacks have experienced a substantial increase, quadrupling in Q4 '21 compared to the previous quarter. These attacks pose a serious threat to the security of IoT devices. They can result in service disruptions, financial losses, and other harmful consequences. Taking proactive measures to prevent and mitigate DDoS attacks is crucial [9]. Therefore, IoT security is critical, and the effects of an IoT attack can be far more devastating than a web attack that temporarily prevents normal user access. Implementing security on IoT devices is difficult due to the heterogeneous and constrained nature of devices [10]. Machine Learning and Deep Learning techniques have proven their effectiveness in detecting and mitigating DDoS attacks over the past few years [11]. In this paper, we discuss the potential of 5G technology to handle the massive volumes of data collected and transmitted by IoT devices. We also highlight the security risks and concerns associated with the rapid spread of these devices. We propose the use of Deep Learning techniques, specifically Convolutional Neural Network (CNN) and Feed forward Neural Network (FNN), to effectively detect and mitigate DDoS attacks in IoT devices within 5G networks. We provide a detailed description of the dataset used in the study, the network infrastructure, and the Deep Learning algorithms used for DDoS detection. Furthermore, a conclusion with a discussion of the results and the potential for future research in this area is provided.

The contribution of this paper includes the following:

- We use the OMNeT ++ simulation tool with the Simu5G framework to generate a dataset for 5G networks involving IoT devices.
- We applied two Deep Learning algorithms, CNN and FNN, using the specifically released dataset for the 5G network, which includes DDoS attacks and normal data traffic.
- We evaluate the performance of models using a confusion matrix.

The rest of the paper is structured as follows: Section II discusses DDoS attacks in IoT. Section III explores the related work on the effectiveness of using Machine Learning and Deep Learning techniques to detect DDoS attacks. In Section IV, we introduce the 5G network, present a novel 5G dataset, and apply Deep Learning models. Section V presents the results. Finally, Section VI concludes the paper, summarizing key points, and suggests potential areas for future work.

## II. DDoS ATTACKS IN IoT

The foundation of IoT is the internet and, as such, it inherits all the potential risks associated with it. Due to the diverse and varied nature of IoT devices, it becomes difficult to ensure the security and privacy of users. DDoS attacks often target IoT devices. Early detection of such attacks can prevent damage and ensure continuity. These attacks can target not only servers, but also network resources, processing units, and storage [12]. The emergence of DDoS attacks was first observed in 1998, but the extent of their destructive capabilities was not fully understood until they were deployed against large companies and organizations in 1999, causing significant damage [13]. Protecting servers and internet services from DDoS attacks is crucial. While traditional DDoS attacks are launched by a network of compromised computers, the use of IoT devices has introduced the more dangerous IoT-DDoS attack. Hackers use hacked IoT device data traffic to overload servers, making it difficult for end-user network hardware to defend against due to high processing power and resource requirements [14].

### *Classification of DDoS attacks in the IoT*

IoT-specific DDoS attacks are not fundamentally different from traditional DDoS attacks, as both rely on exploiting vulnerabilities to overload systems. However, the wide range of IoT devices in existence leads to greater diversity and complexity in these types of attacks. As shown in Figure1, it is possible to categorize these assaults into three groups according to the tactics used by the attackers [15].

1) **Application layer attacks:** involve attempting to breach the application layer of the network infrastructure. This can occur when packets are dropped, as a result of overwhelming the application or web server with a flood of HTTP(Get/Post) requests and other requests that target system software such as Windows, Apache, OpenBSD, and others [15].

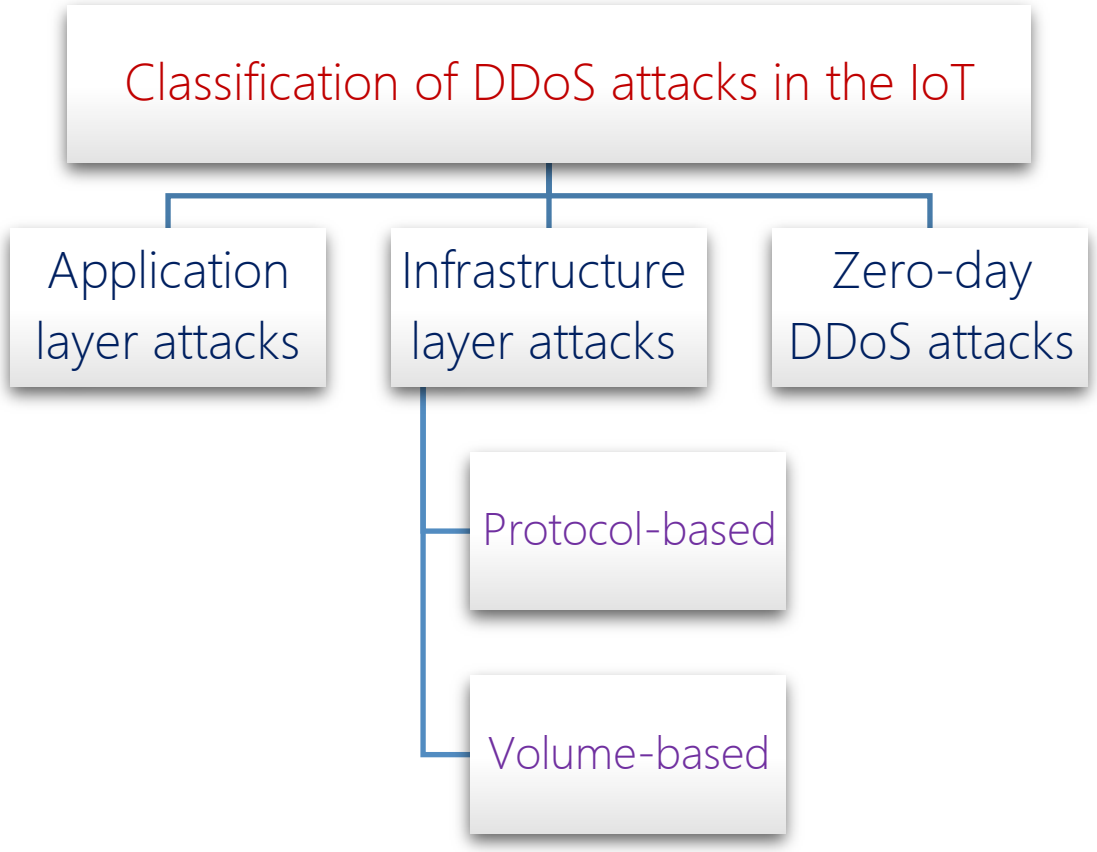

**Fig.1.** Classification of DDoS attacks in IoT *[15]*.

2) **Infrastructure layer attacks:** attacks targeting the infrastructure layer of IoT systems aim to make the target system inaccessible by exploiting vulnerabilities in the transport or network layers. They come in two forms: protocol-based and volume-based attacks. Common tactics used include reflection and amplification, where the attacker manipulates IP addresses to redirect traffic back to the target, creating network congestion. Amplification amplifies the size of the response to a smaller request, leading to significant bandwidth wastage [15].
   - **Protocol-based:** also known as Resource Depletion attacks, are designed to consume the resources of servers and other communication devices like firewalls and load balancers. These attacks are quantified by the number of packets they send per second (Pps), examples include SYN floods [15].
   - **Volume-based:** also known as Bandwidth Depletion attacks, are characterized by overwhelming the target system's available bandwidth with an excessive amount of data measured in bits per second (Bps). They employ amplification and reflection techniques, making them relatively easy to execute. Examples of attacks in this category include UDP/TCP floods and ICMP floods [15].

3) **Zero-day DDoS attacks:** refer to a new category of unknown or novel DDoS attacks that exploit unknown vulnerabilities in systems. This type of attack has gained popularity among cybercriminals in recent times [15].

## III. RELATED WORK

Machine learning and Deep Learning techniques have proven effective in detecting DDoS attacks on IoT devices. Many researchers in this field have conducted many experiments and research projects. Doshi et al. [16] developed a Machine Learning pipeline that performs data collection, feature extraction, and binary classification for DDoS detection of IoT traffic DDoS detection. They used Random Forest (DF), K-Nearest-Neighbor, Support Vector Machine, Decision Tree, and Neural network. Their results showed excellent performance with an overall detection accuracy of 99\%. Ma et al. [17] developed a novel CNN model to detect DDoS attacks on IoT. The proposed model achieved an accuracy of 92% compared to the classical CNN, which reached 89%. Alnuman et al. [18] used OMNeT++ to simulate an IoT network in the home, including a DDoS attack. Regular and attack-injected traffic is created to test the accuracy of Machine Learning methods for detecting DDoS attacks in IoT networks. They analyzed the generated dataset by using DF, Decision Jungle, and a Boosted Decision Tree. The accuracy of DF was 83.80%, the Decision Jungle was 83.20%, and the Boosted Decision Tree was 99.90%.

## IV. METHODOLOGY

In this section, we will discuss the implementation of our 5G network, extracting a dataset from it, and applying deep learning methods to analyze the data.

### *1. 5G Dataset*

To establish our 5G network, we chose OMNeT++ for several reasons. It offers a user-friendly graphical interface (GUI) that simplifies simulation management. Furthermore, OMNeT++ is an open-source tool [19]. OMNeT++ integrates the Simu5G framework, which significantly enhances the efficiency of creating 5G networks. We installed OMNeT++ on an Ubuntu operating system, effectively integrating it with both the Simu5G and INET frameworks. Moreover, we utilized the NED programming language, and the respective versions of these components are presented in Table1.

**Table .1** Experiment Tools for 5G network

| No | Tool | Version |
|---|---|---|
| 1 | Ubuntu OS | 20.4 |
| 2 | OMNeT++ | 6.0.1 |
| 3 | Simu5G | 1.2.1 |
| 4 | INET | 4.5 |

In this study, we present the network architecture of our 5G network, as shown in Figure2. This architecture includes critical network components, such as a gNodeB, backgroundCell, router, 100 New Radio User equipment's (NRUe), three hosts, and components of 5G Core (5GC).

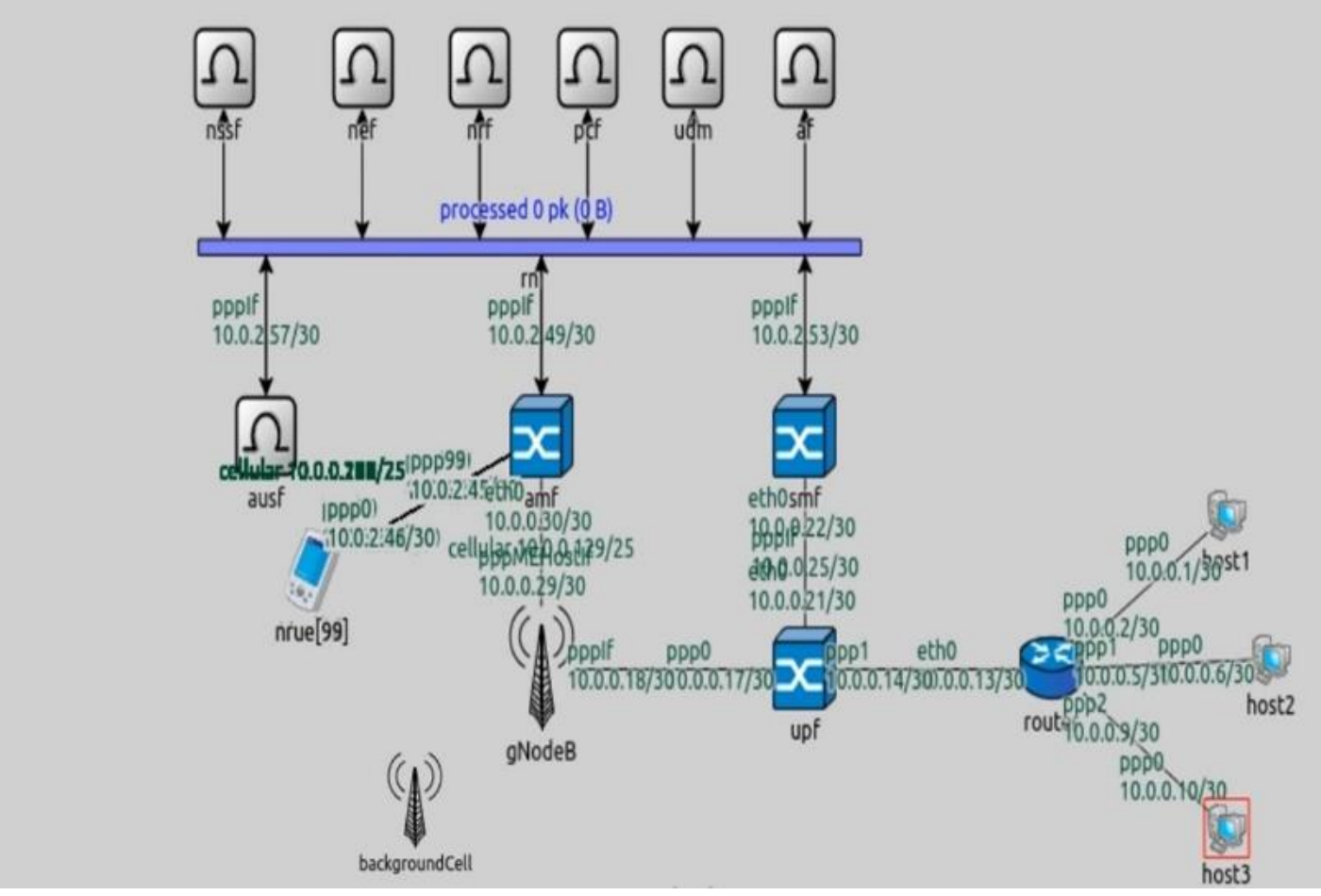

**Fig.1.** Architecture of 5G network.

### *1.1 Details of the simulation*

In our network configuration process, we have dispersed the locations of the individual nodes throughout the network topology. This strategic placement of network nodes enables efficient communication and supports the effective distribution of network resources. Furthermore, we have customized specific settings to suit our network requirements, which are outlined below:

1) **Ipv4NetworkConfigurator:** we configured it to enable the dumping of various network configuration details during the simulation. These details included the IP addresses assigned to each network interface in the topology, network topology information such as which nodes were connected to which others, bandwidth and delay information, and routing information such as which paths traffic took between nodes. By enabling these settings, the simulation was able to output detailed information about the network configuration and behavior, which could be utilized to debug issues or analyze the simulation results.
2) **Routing settings:** we configure the routing settings by assigning the Global Address Resolution Protocol (ARP) to all nodes with an IPv4 module and how packets are routed between nodes. The ARP protocol is essential for enabling communication between devices in an IP over Ethernet network, as it enables efficient mapping of IP addresses to MAC addresses [20].
3) **Visualization settings:** the purpose of the visualization settings is to display the IP addresses on the simulation image.
4) **General Physical Layer parameters:** the parameters required in the configuration of the NRUe devices and the gNodeB used.

### *1.2 Network Scenarios*

1) **First scenario Normal data traffic:**
   As shown in Figure3 our network operates similarly to other networks, allowing for the natural flow of data traffic between devices. All devices connected to the 5G network can communicate with each other by sending PING, ensuring that data traffic reaches its intended destination.
2) **Second scenario DDoS attack:**
   As previously discussed, all devices in the network can communicate and transmit with each other through 5G networks. However, in this scenario, we have introduced three hosts devices that are directly connected to the router, as depicted in Figure 3. Our objective is to generate malicious traffic in the form of a DDoS attack with the aim of disrupting communication between all nodes. We achieve this by generating packets of unusually large size (1000 bytes) from all hosts and transmitting them at an exceptionally high speed (0.001 seconds) continuously. It is worth noting that this packet size is considerably larger than the largest packet size that a host is typically required to accept, which is only 576 bytes [21]. The DDoS attack propagates to all units on the network, resulting in a disrupted communication pattern that is considered abnormal traffic. Consequently, communication breaks down, and the devices cannot communicate with each other.

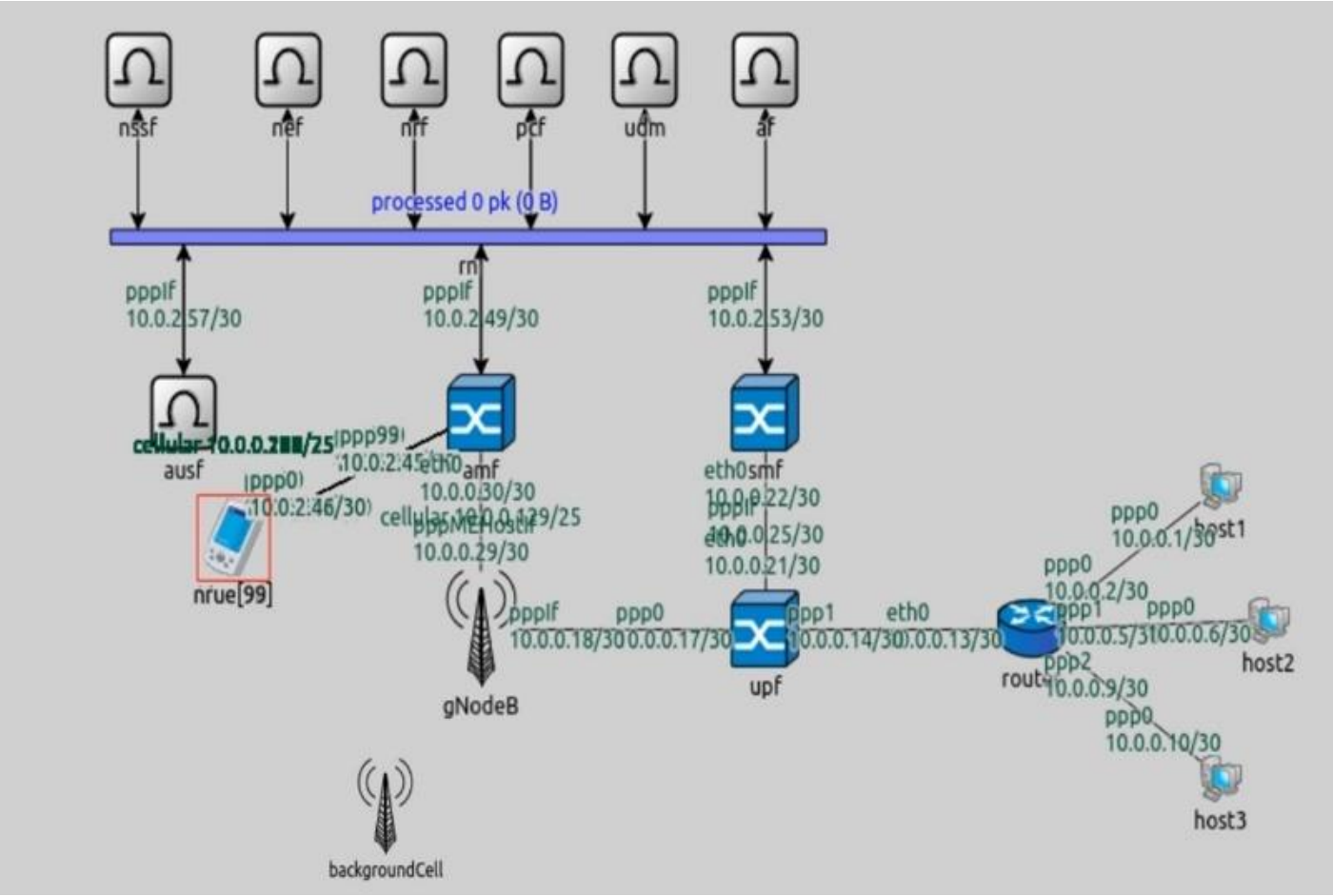

**Fig.2.** First scenario Normal data traffic.

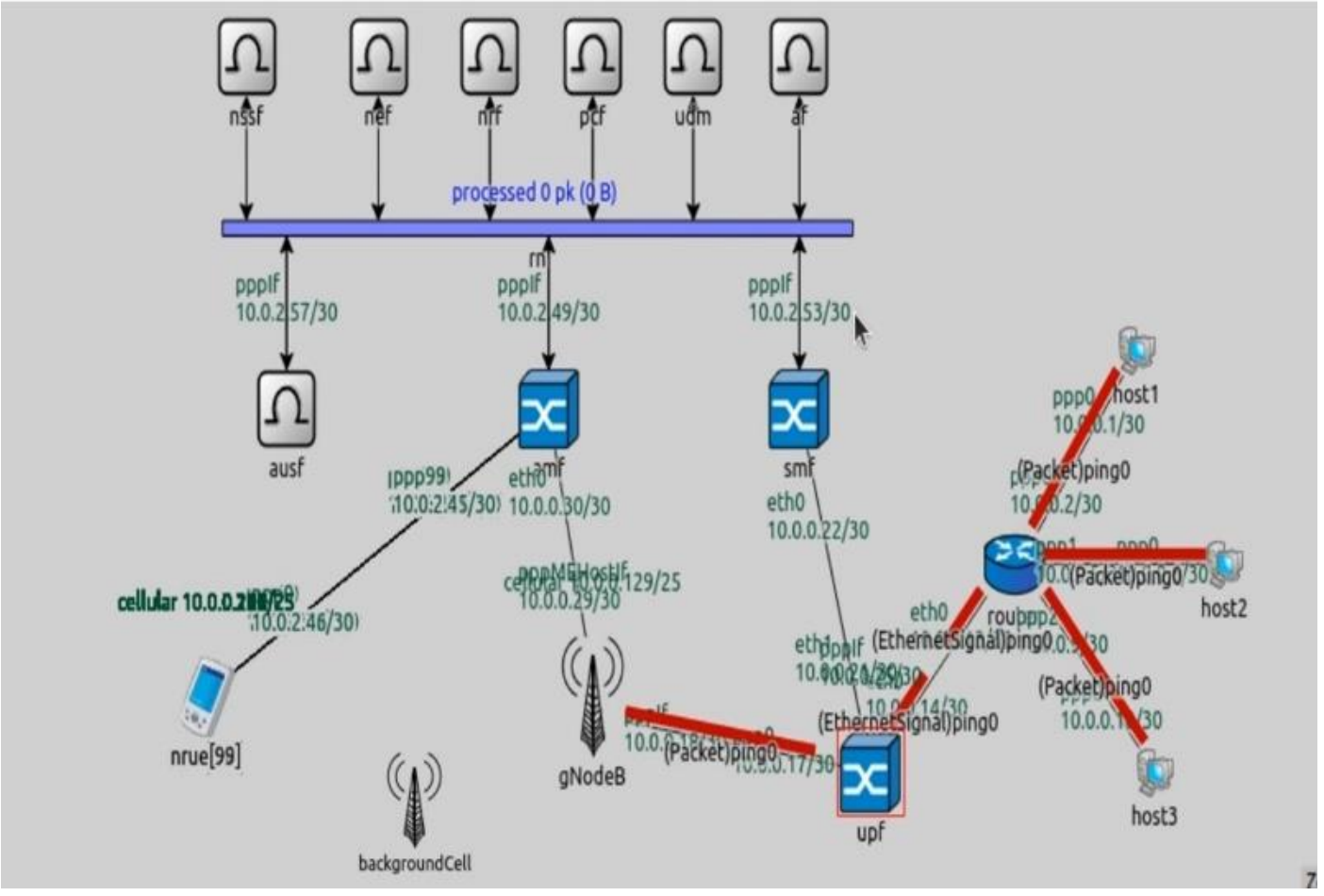

**Fig.3.** Second scenario DDoS attack.

### 2. *Creating a Dataset with 5G Networks*

We compiled a dataset that consists of 512,666 samples, including both benign data and DDoS attacks, with 16 features shown in the following Table 2. The count of benign samples reached 256,354, while the count of DDoS attacks amounted to 256,312.

**Table.2** Features in 5G network dataset

| No | Features | No | Features |
|---|---|---|---|
| 1 | sumweights | 9 | mean |
| 2 | type | 10 | stddev |
| 3 | module | 11 | min |
| 4 | name | 12 | max |
| 5 | attrname | 13 | underflows |
| 6 | attrvalue | 14 | overflows |
| 7 | value | 15 | binedges |
| 8 | count | 16 | binvalues |

### 3. *Preprocessing*

Our dataset contains a significant number of null values, and we need to remove them. We will remove the columns with the least amount of data, specifically 'count', 'sumweights', 'mean', 'stddev', 'min', 'max', 'underflows', 'overflows', 'binedges', and 'binvalues'. We filling missing values using the forward-fill method and converting columns to numeric values.

### 4. *Applying Deep Learning*

We applied two Deep Learning models: CNN and FNN.

1) **CNN:** The CNN model has three 1D convolutional layers and three 1D MaxPooling layers. The first layer has 64 filters and a kernel size of 8, while the second and third layers have 32 and 16 filters with kernel sizes of 16 and 3, respectively, followed by a dropout layer that randomly drops out half of the input units during training to prevent overfitting. We opted for a 1D convolutional layer over a 2D convolutional layer due to its faster training time and the absence of a requirement for a dedicated Graphics Processing Unit (GPU) [22]. The output of the last layer is flattened and passes through two fully connected dense layers with ReLU and sigmoid activation functions, respectively.
2) **FNN:** The neural network contains three hidden fully connected layers. Each hidden layer performs a set of computations on the input data to capture relevant patterns and representations. The first hidden layer consists of 64 units. The second hidden layer with 32 units. The third hidden layer, containing a single unit. The output of the final layer uses a sigmoid activation function.

## V. Results and Discussion

To evaluate the effectiveness of our models in the detection of DDoS attacks, we use a variety of measures derived from established equations. These metrics are essential for evaluating our models.

a) Accuracy: Calculating the accuracy rate for the entire model is according to the following equation:

$$\frac{TP+TN}{TP+TN+FP+FN} \tag{23}$$

b) Detection Rate: Calculating the detection rate for the entire model is according to the following equation:

$$\frac{TP}{TP+FN} \tag{23}$$

c) False alarms: Calculating the false alarms for the entire model is according to the following equation:

$$\frac{FP}{FP+TN} \tag{24}$$

whereas:
TP = True Positives.
TN = True Negatives.
FP = False Positives.
FN = False Negatives.

The dataset is divided into two sets: 80% for training and validation, and 20% for testing. Within the training and validation set, 70% is allocated for training data, and 10% for validation data. It's essential to recognize that our dataset contains features with different value ranges. To maintain a balanced classification process and avoid the potential dominance of larger values over smaller ones, we use Min-Max

Scaler. The Min-Max Scaler, which employs linear normalization to ensure that all feature values are situated within the (0,1) range [25] [26] [27]. In this study, we conduct multiple experiments to fine-tune the following hyperparameters: the learning rate and kernel size. It's worth noting that in all these experiments, we keep the number of epochs constant at 10, and learning rate 0.001. The results of the models are presented in Table 3.

**Table.1** Result of applying Deep Learning methods to the 5G dataset

| Model | Accuracy | Precision | Recal | F1 Score |
|---|---|---|---|---|
| CNN | 99.74% | 99.87% | 99.61% | 99.74% |
| FNN | 99.53% | 99.53% | 99.54% | 99.53% |

The charts below illustrate a comparison between the accuracy, precision, recall and F1 score of the Deep Learning models that are used.

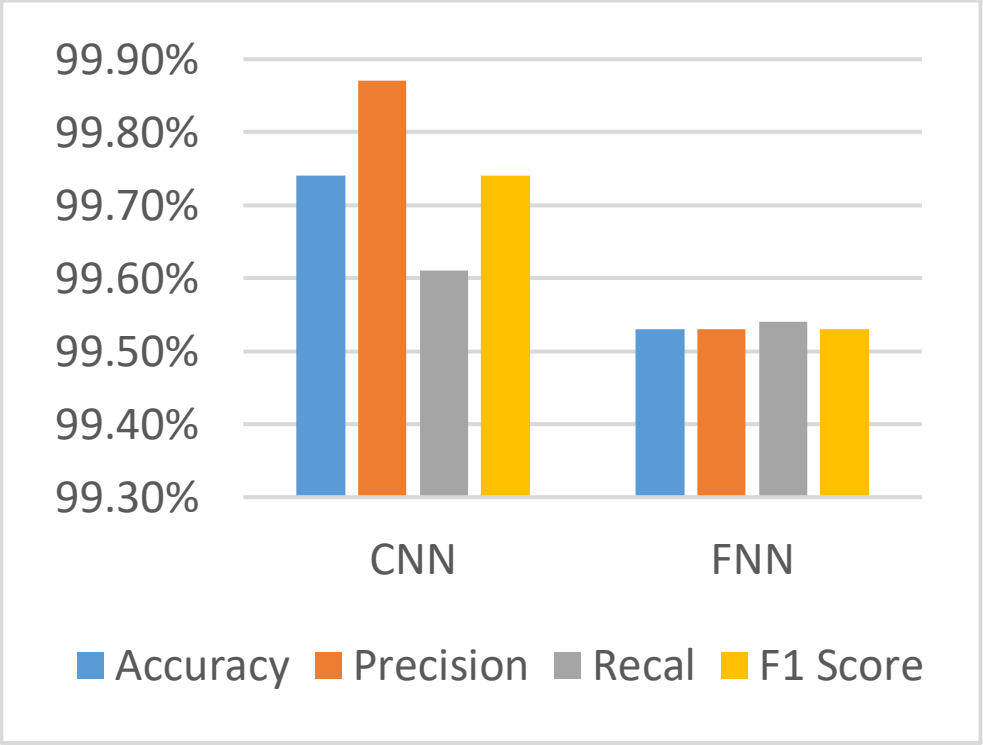

**Fig.4.** Results of Experiment.

All our models deliver exceptional performance, boasting an accuracy and recall rate of 99%.

## VI. Conclusions

In this paper, we present a deep learning-based framework to detect DDoS attacks in 5G networks for IoT devices. We implemented CNN and FNN models to identify DdoS attacks in a dataset specifically designed for 5G networks and Internet of Things devices. We evaluate the performance of our models using a confusion matrix. The results clearly indicate that these models perform exceptionally, consistently achieving more than 99% in all performance metrics.

In future work, we can try different models apart from those used in our study. We also recommend using the NETA framework within OMNeT++. This framework enables you to generate various attacks and detect them using the integrated TensorFlow framework within OMNeT++. Moreover, we suggest creating complex scenarios using the OMNeT++ frameworks. For example, a group of drones that cover various locations. These drones can be moved using the X-Plane software. These drones can potentially be exploited to generate DdoS attacks. To effectively detect and mitigate these attacks, we suggest using the TensorFlow framework integrated within OMNeT++.

## VII. Acknowledgment

Words could not express our gratitude to Dr. Abdullah Alomari and Dr. Najib Ben Aoun, graciously shared their knowledge, and provided invaluable feedback that enabled us to refine this study.